\documentclass{appolb}
\usepackage[T1]{fontenc}
\usepackage{epsfig}
\usepackage[utf8]{inputenc}
\usepackage{longtable}

\begin{document}
\author{Cezary Juszczak
\address{Institute for Theoretical Physics\\University of Wrocław}
}
\title{Running NuWro
\thanks{Presented by C.\ Juszczak at the 45th Winter School in
Theoretical Physics ``Neutrino Interactions: from Theory to Monte
Carlo Simulations'', L\k{a}dek-Zdr\'oj, Poland, February 2--11, 2009.}%
}
\def\MeV{{\rm MeV}}

\maketitle
\begin{abstract}
 The NuWro Neutrino Event Generator developed by the Wrocław Neutrino
 Group (WNG) is lightweight but full featured. It handles all interaction
 types important in neutrino-nucleus scattering as well as
 DIS hadronization and intranuclear cascade.
Its input file, by default {\tt params.txt}, is a plain text file and
the output file, by default {\tt eventsout.root},
is a {\tt root}\footnote{{\tt root} and {\tt rootcint}
are parts of the CERN software http://root.cern.ch/ and {\tt .root}
is the file extension of data files used and produced by the software.} file
which can be analyzed by means of the included \verb`myroot` program,
 or by standard \verb`root`, after loading supplied dictionary library {\tt event1.so}.
%~ should be loaded before reading the file.
%~ In more complicated cases it is possible to add desired methods to the
%~ {\tt event} class or to write a {\tt root} based  \verb`C++` program
%~ to perform the analysis.
\end{abstract}

\section{Installing NuWro}

NuWro is a neutrino event generator developed by the Wrocław Neutrino Group.
It can be downloaded as a tar ball from:\\[3mm]
%\begin{verbatim}
%http://borg.ift.uni.wroc.pl/websvn
%\end{verbatim}
{\tt http://borg.ift.uni.wroc.pl/websvn}\\[3mm]
Alternatively, the subversion command:
\begin{verbatim}
svn export svn://borg.ift.uni.wroc.pl/pub/nuwro
\end{verbatim}
can be used to create a directory {\tt nuwro} containing the copy of the
 current NuWro sources.
Then it should be enough to type:
\begin{verbatim}
cd nuwro
make
\end{verbatim}
to build the program, provided the {\tt root}
 software configured with the Pythia6 library is installed on one's computer.

%~ To run the program in the default mode type:
%~ \begin{verbatim}
%~ ./nuwro
%~ \end{verbatim}
%~ and to use non-default input and output type:
%~ \begin{verbatim}
%~ ./nuwro -i myinput.txt -o myoutput.root
%~ \end{verbatim}
%~ Any number of  values of the parameters defined in the input file may
%~ be overwritten by values
%~ specified on the command line as follows:
%~ \begin{verbatim}
%~ ./nuwro -p "parname1=parvalue1" -p "parname2=parvalue2" ...
%~ \end{verbatim}

\subsection{Installing root with Pythia6}
Unfortunately the {\tt libPythia6.so} library is not included in the
 {\tt root} distribution.
It must be downloaded and built separately before building root.
It is best done by
typing either\\ {\tt build\_pythia6.sh\footnote{{\tt build\_pythia6.sh}
is a script by Robert Hatcher <rhatcher@fnal.gov> which can be downloaded
from {\tt http://home.fnal.gov/$\sim$rhatcher/build\_pythia6.sh.txt}}  gfortran}\\
or\\ {\tt build\_pythia6.sh g77}\\ depending on which fortran compiler you have.

 The resulting {\tt libPythia6.so} file should be placed in the {\tt lib}
 directory of the {\tt root} source
 tree\footnote{Obtain root sources
  from {\tt http://root.cern.ch/drupal/content/downloading-root}}. Then
the root software should be configured and build with the following commands
\footnote{Several libraries are needed to build root and Pythia6.
Under Ubuntu 9.04 the following packages must be installed:
{\tt g++, gfortran, libX11-dev, libxft-dev, x11proto-xext-dev, libXpm-dev,
libXext-dev.}} issued in the {\tt root} sources directory:
\begin{verbatim}
 ./configure --with-pythia6-libdir=`pwd`/lib
 make
\end{verbatim}
To be able to run {\tt root} from any location and compile {\tt root} based programs like {\tt nuwro},
the following lines should be added to one's {\tt .bash\_profile}\footnote{For a system wide installation,
a file {\tt root.sh} consisting of these three lines should be placed in {\tt /etc/profile.d/} directory, instead.
Under Ubuntu it is convenient to put them to {\tt .bashrc}}
\begin{verbatim}
export ROOTSYS= path to directory where you made root
export PATH=$PATH:$ROOTSYS/bin
export LD_LIBRARY_PATH=$LD_LIBRARY_PATH:$ROOTSYS/lib
\end{verbatim}

\section{Running Nuwro}
To run the program in the default mode type:
\begin{verbatim}
./nuwro
\end{verbatim}
and to use non-default input and output type:
\begin{verbatim}
./nuwro -i myinput.txt -o myoutput.root
\end{verbatim}
Any number of values of
 the parameters defined in the input file may be overwritten by values
specified on the command line as follows:
\begin{verbatim}
./nuwro -p "parname1=parvalue1" -p "parname2=parvalue2" ...
\end{verbatim}
which is very useful when running {\tt nuwro} in batch mode with
changing parameter values and  output locations.

\subsection*{NuWro input file}
All the parameters of the NuWro generator are read from a file, usually {\tt params.txt}.
The structure of this file is quite simple --
each line is either a comment (if it begins with {\tt \#}):
\begin{verbatim}
# This is an example of a comment line
\end{verbatim}
or a substitution:
\begin{verbatim}
parameter_name = parameter_value
\end{verbatim}
Parameter names coincide with the
corresponding {\tt C++} variable names inside the program.
Only four parameter types are used: {\tt int} (integer number),
 {\tt double} (floating point number),
 {\tt vec} (3D vector initialized by three white space separated numbers) and
 {\tt string} (stretches to the end of line).

The meaning of the parameters is briefly explained and their possible
values listed in the commented lines of the {\tt params.txt} file itself.
Let us summarize the meaning of the most important the parameters here.

\subsection*{Basic parameters}
\noindent The test events are not stored in the output file,
but the average of their weights becomes the total cross section in each channel:
\begin{verbatim}
number_of_test_events = 1000000
\end{verbatim}
The  number of unweighted events to be stored in the output is given by:
\begin{verbatim}
number_of_events = 500000
\end{verbatim}

%\subsection*{The random number generator behaviour}
%~ \noindent The random number generator seed is determined from the parameter:
%~ \begin{verbatim}
%~ random_seed = 1
%~ \end{verbatim}
%~ with the following logic: {\tt 0} - use \verb`time(NULL)` as the seed,
%~ {\tt 1} - read seed from the file {\tt random\_seed} and save seed in the
 %~ file when program completes.
%~ Other values are used directly as the random seed, and should be avoided.
%#######################################################################
\subsection*{The beam definition}
\noindent
At present, only beams of identical neutrinos flying in the same
direction are allowed. The beam direction coordinates $x$, $y$, $z$ and
neutrino PDG code must be specified e.g.
\begin{verbatim}
beam_direction     = 0 0 1
beam_particle      = 14
\end{verbatim}
\noindent The neutrino energy given in MeV can be either fixed e.g.
\begin{verbatim}
beam_energy = 1000
\end{verbatim}
or given as a histogram encoded in the sequence of numbers: $E_{min}$,
$E_{max}$, $n_1$, $n_2$, $n_3$ \ldots, $n_k$.
\begin{verbatim}
beam_energy = 1000 4000 1 2 3
\end{verbatim}
The number of beans $k$ is inferred from the length of the sequence and
the bean width is $(E_{max}-E_{min})/k$.
Definitions of a few popular beams are included in the {\tt params.txt}
file and it is enough to uncomment
the corresponding (sometimes very long) line to use one of them.

\subsection*{The target nucleus definition}
\noindent The target nucleus is defined by the following parameters:
\begin{verbatim}
nucleus_p       = 8   // number of protons
nucleus_n       = 8   // number of neutrons
nucleus_density = 1   // 1 - constant, 2 - realistic density
\end{verbatim}
and the switch identifying the nucleus model to be used:
\begin{verbatim}
nucleus_target		= 1
\end{verbatim}
with the following allowed values:
{\bf 0} - free nucleon; {\bf 1} - Fermi gas; {\bf 2} - local Fermi gas;
{\bf 3} - FG with Bodek-Ritchie momentum distribution; {\bf 4} - "effective" spectral function (carbon or oxygen);
{\bf 5} - deuterium.
%{\bf 6} - deuterium with constant binding energy equal \verb`nucleus_E_b`.
%Nuclear density model used in the cascade code:
%\begin{verbatim}
%nucleus_model       = 1
%\end{verbatim}
%has two possible values: {\bf 0} - constant nuclear density,
%{\bf 1} - realistic nuclear density profile from \cite{}.
%#######################################################################

In cases where the Fermi Gas model is used it is possible to specify:
 \begin{verbatim}
nucleus_E_b      =  27 // nucleon bounding Energy in MeV
nucleus_kf       = 225 // Fermi momentum in MeV
\end{verbatim}

\subsection*{The physics effects switches}
\noindent Nonzero values of the switches:\\[2mm]
\verb`dyn_qel_cc`,
\verb`dyn_res_cc`,
\verb`dyn_dis_cc`,
\verb`dyn_coh_cc`,\\
\verb`dyn_qel_nc`,
\verb`dyn_res_nc`,
\verb`dyn_dis_nc`,
\verb`dyn_coh_nc`\\[2mm]
%~ \begin{verbatim}
%~ dyn_qel_cc = 1
%~ dyn_qel_nc = 1
%~ dyn_res_cc = 1
%~ dyn_res_nc = 1
%~ dyn_dis_cc = 1
%~ dyn_dis_nc = 1
%~ dyn_coh_cc = 1
%~ dyn_coh_nc = 1
%~ \end{verbatim}
indicate that
quasi elastic, resonant, deep inelastic, and coherent events should be
generated. There are {\tt nc}/{\tt cc} variants to separately control generation of
neural current and charge current events.\\[2mm]
The quasi-elastic cross sections depend on the values of axial masses:
\begin{verbatim}
qel_nc_axial_mass= 1030   //MeV
qel_cc_axial_mass= 1100   //MeV
\end{verbatim}
and the choice of the form factors:
\begin{verbatim}
qel_cc_vector_ff_set = 1   // only 1 is possible
qel_cc_axial_ff_set = 1    // 1 - dipole form factors,
                           // 2, 3, 4 - two-fold parabolic modifications
\end{verbatim}
%~ There is only one set of the vector form factors at the moment and
%~ four sets of axial form factors:
%~ {\bf 1} - dipole, {\bf 2, 3, 4} - two-fold parabolic modification
%~ with different parameters.
By specifying a nonzeoro value of the parameters:
\begin{verbatim}
flux_correction = 1
qel_relat       = 1
\end{verbatim}
the difference in neutrino flux between nucleon and nucleus frames
is accounted for.

\noindent Finally, running the cascade code with
Pauli-blocking of the intermediate nucleons,
is achieved by the lines:
\begin{verbatim}
kaskada_on 	 = 1
pauli_blocking   = 1
\end{verbatim}
%kaskada_debug 	 = 0
Some parameters are not listed here because they are used only for testing
{\tt nuwro} and setting them to nondefault values would result in obtaining
 incorrect  results.

\section{Analysing the output}
The output of {\tt nuwro} is a {\tt root} file, usually {\tt eventsout.root}.
This file contains only one object {\tt treeout} which is a {\tt TTree}
with a single branch {\tt e} containing the {\tt event} objects.
The easiest way to access this file is by means of the {\tt myroot} program:
\begin{verbatim}
./myroot eventsout.root
\end{verbatim}
which is build together with {\tt nuwro}.
It is a version of {\tt root} containing a dictionary
for the class {\tt event} and all its dependencies.
%In the {\tt nuwro} output analysis it is important to know that
Each {\tt event} contains the following data:
\begin{description}
\item[{\tt params}] - parameters read from the input file and the command line.
\item[{\tt flags}] - set of booleans ({\tt coh, qel, dis, res, nc, cc, anty})
to easily filter events based on primary vertex interaction type.
\item[{\tt dyn}] - integer identifying the dynamics used in the primary vertex.
\item[{\tt in, tmp, out, post, all}] - are STL\footnote{STL
- stands for the C++ Standard Template Library} vectors of {\tt particles}
(incoming, temporary, outgoing, post, and all the particles including
the intermediate cascade particles).
\item [{\tt weight}] - a number proportional to the cross section.
\end{description}
It has also a number of useful methods: {\tt q2(), s(), n(), W(), nof()}, etc.

A {\tt particle} is a Lorentz fourvector with coordinates
{\tt t, x, y, z} denoting its energy
and momentum components, supplied with is pdg code, mass, position
fourvector {\tt r} (used only by the cascade code)
and some less important attributes. Several methods are added with self
explanatory names like {\tt E(), Ek(),
momentum(), v(),} \ldots.

By convention {\tt in[0]} is always the beam particle and
 {\tt out[0]} is the outgoing lepton.
Thanks to the way the {\tt root} interpreter handles STL vectors,
that you can type {\tt out[0].t} for the outgoing lepton energy,
 {\tt out.Ek()} for the energy of all outgoing particles and
 {\tt @out.size()} for the number of outgoing particles
 (here {\tt out[0]} stands for the first outgoing particle, {\tt out} for any outgoing particle and {\tt @out}
 for the whole vector of outgoing particles).

The possibility to use
native C++ methods from the root interpreter significantly simplifies
the analysis.
 It is possible to obtain many interesting plots
with just one command:\\[2mm]
{\tt treeout->Draw("out.mass()");}
//masses of outgoing particles.\\[2mm]
%{\tt treeout->Draw("@out.size()"); } or \\
{\tt treeout->Draw("n()"); }
  // number of outgoing particles \\[2mm]
{\tt treeout->Draw("nof(111)","flag.dis*flag.cc");}\\
  // number of $\pi^0$ produced in deep inelastic charge current events\\[2mm]
{\tt treeout->Draw("out.momentum()","out.pdg==111");}\\
  // momenta of outgoing $\pi^0$. \\[2mm]
{\tt treeout->Draw("all.r.x:all.r.y:all.r.z"); } \\
  // places where the interactions took place in the cascade code\\[2mm]
{\tt treeout->Draw("q2()","flag.dis");}\\
  // proportional to $d\sigma/dq^2$ differential cross section in the DIS channel

In more complicated	cases it is possible to write a script or a
root based C++ program to perform the analysis.
It is also possible to add more methods to the {\tt particle}
and {\tt event} classes.
More information on NuWro can be found
on {\tt http://wng.ift.uni.wroc.pl/nuwro}.

\end{document}